# Size Adaptive Region Based Huffman Compression Technique

Utpal Nandi[#1], Jyotsna Kumar Mandal[#2]

[#1]Dept. of Computer Sc. & Engineering, Bankura Unnayani Institute of Engineering
Bankura-722146, West Bengal, India

[#2] Dept. of Computer Sc. & Engineering, University of Kalyani, Nadia –741235,
West Bengal, India

[1]nandi_9utpal@yahoo.com

[2]jkm.cse@gmail.com

*Abstract— A loss-less compression technique is proposed which uses a variable length Region formation technique to divide the input file into a number of variable length regions. Huffman codes are obtained for entire file after formation of regions. Symbols of each region are compressed one by one. Comparisons are made among proposed technique, Region Based Huffman compression technique and classical Huffman technique. The proposed technique offers better compression ratio for some files than other two.*

## I. INTRODUCTION

The data compression techniques reduce the amount of data required to represent a source of information to reduce the storage space requirement and the time of data transmission over network. There are two major class of data compression techniques i.e. loss-less [1-6] and lossy [6]. The loss-less techniques generate exact duplicate of the original data after compress/expand cycle. But, the lossy techniques concede a certain loss of accuracy. One of the well established loss-less technique is Huffman Coding [4,6] which is based on the frequency of elements of entire file. If an element has maximum frequency, it gets shortest code. But if we divide a file into a number of regions, it is obvious that in each region the maximum frequency element may not the maximum frequency element of entire file and has large code length. If the large codes produced by Huffman coding are used for the elements which has maximum frequency for each region ,the size of compressed file increases. In light of this Region Based Huffman (RBH) [1] coding has been introduced. The RBH coding technique divides the total input file/stream into a number of regions N. The maximum frequency elements for each region are calculated. Huffman codes are obtained based on frequency of elements for entire file/stream. Now for first region, if the code length of maximum frequency element of that region is larger than the code length of maximum frequency element of entire file/stream, the code between maximum frequency element of that region and maximum frequency element of entire file/stream is interchanged. This interchanged information is attached with the compressed file/stream. The elements of that region are compressed with the changed codes and interchanged codes are reset. Otherwise, same symbol code table is used. Similarly, all other regions are compressed repeatedly. The main problem of RBH coding is that the performance depends on the number of regions of the file and therefore also on the size of region of the file. Compression ratios of same file with different region size are not same. It is very difficult to determine the optimum region size that offers maximum compression of a file. Because, it depends on the symbols of the file. As different region contains different frequency of symbols, the compression ratios of different region are also not same. For example, let us Consider a file/stream containing the message CACBABCBCCABACBA BABACBBADBDBEB (say MSG). For the same file/stream (MSG), compression ratio is 47.9**%** for region size 10 and compression ratio is 45% for region size 6. Therefore, the proper region size (or number of region) must be chosen for better compression of file/stream. Modified Region Based Huffman (MRBH) [1] coding also suffers from the same problem if the optimum value of number of region does not lie in the specified range. Fixed size regions are not able to adapt its size based on symbols that offers better compression. To overcome the limitation of finding optimum region size of fixed size region based compression techniques, a technique is proposed which has the ability to adapt its region size based on symbols and termed as Size Adaptive Region Based Huffman (SARBH) coding. The proposed technique has been discussed in section II. Results have been given in section III and conclusions are drawn in section IV.

## II. THE PROPOSED TECHNIQUE

Most of the time, it is found that the sequence of character's ASCII values in the file are adjacent ASCII values. That is ASCII value differences among adjacent characters are not so high. The proposed algorithm uses this concept. Our aim is to group sequence of characters into regions such that the differences among the ASCII values of characters in a region do not exceed a specified value (r). Therefore, after grouping into regions, the information of each region can be preserved by storing the number of symbols, minimum ASCII value and the differences among other ASCII values of symbols in the region with the minimum ASCII value. After that each region contains ASCII values not exceeding the specified value except first two (the number of symbols and the minimum ASCII value). In this way, variable length regions ( R1 , R2 , R3 . . . Rn) of input file / stream using with a specified value(r) is formed. The frequencies of all the symbols of the input file / stream whose ASCII value lies in the range 0 to r-1 are obtained. Huffman Codes of all symbols whose ASCII value lies in the range 0 to r-1 of input file / stream is also calculated. During compression, for each region first two symbols (number of element and minimum ASCII value symbol) are kept unchanged and all other symbols (whose ASCII values lie in the range 0 to r-1) are coded by corresponding Huffman code.

For example, let us consider a file/stream containing the message – **ABAABDADAAWXXZXWXY ZXXYPQPSQS PR** (say MSG1). MSG1 is grouped into a number of regions with specified value(r) =16 in such a way that each region does not have two symbols with ASCII value difference greater than or equal to 16 as shown below in Fig. 1.

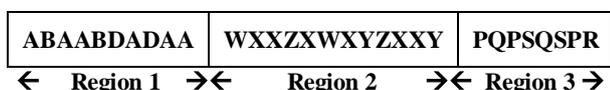

Fig. 1: Symbols of variable length regions

Information of each regions are kept by storing the number of symbol of each region, minimum ASCII value of all the symbols and ASCII value difference of all the symbols with minimum ASCII value of symbol in the corresponding region as shown below in Fig. 2.

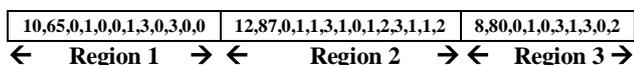

Fig. 2: variable length regions of MSG1

After formation of regions, frequencies of all the numbers in the range 0 to 15 are found as given in TABLE I and Huffman tree based on the frequency of numbers are obtained as shown in Fig. 3. Code of each numbers are thus obtained as given in symbol code table TABLE II.

TABLE I : FT of Symbols

| Number | 0 | 1 | 2 | 3 |
|---|---|---|---|---|
| **Frequency** | 11 | 10 | 3 | 6 |

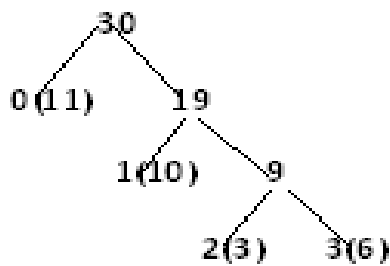

Fig. 3 : Huffman tree based on Table I

Table II : SCT

| Number | 0 | 1 | 2 | 3 |
|---|---|---|---|---|
| code | 0 | 10 | 110 | 111 |

During compression, for each region first two symbols are kept unchanged and all other symbols are coded by corresponding Huffman code. Therefore, the compressed message for entire file/message stream will be-**10,65,0,10,0, 0, 10,111,0,111,0,0,2,87,0,10,10,111,10,0,10,110,111,10,10,110, 8,80,0,10,0,111,10,111,0,110.** The calculation of compression ratio may be done as follows: Original message size = 30x8 bits = 180 bits, Only Compressed message size(excluding first two numbers) = 58 bits ,Frequency Table size = 6x8 bits = 48 bits, size of first two numbers of three region=3x2x8 bits = 48bits, Total Compressed message size = ( 58 + 48+ 48 ) bits = 154 bits , Compression ratio = { ( 240 – 154 ) / 240 }X100 % = 35.83 %.

### III. RESULTS

Comparison of compression ratios among Huffman technique, MRBH coding with range 10 - 25 and proposed **SARBH** coding have been made using five different type of files as shown in TABLE 3 with specified value(r) as 32. Graphical representation of the same is shown in Fig. 4.

TABLE III: Comparison of compression ratios in different techniques

| File Name | %Compression | | |
|---|---|---|---|
| | Classical Huffman | MRBH with range 10 - 25 | Proposed SARBH |
| Circle.java | 35.48 | 35.50 | 35.80 |
| Selection.exe | 19.47 | 19.86 | 19.82 |
| Dolly.doc | 39.12 | 40.13 | 40.10 |
| Chpst.dll | 24.42 | 24.81 | 24.80 |
| Dummy.txt | 34.60 | 34.59 | 35.01 |

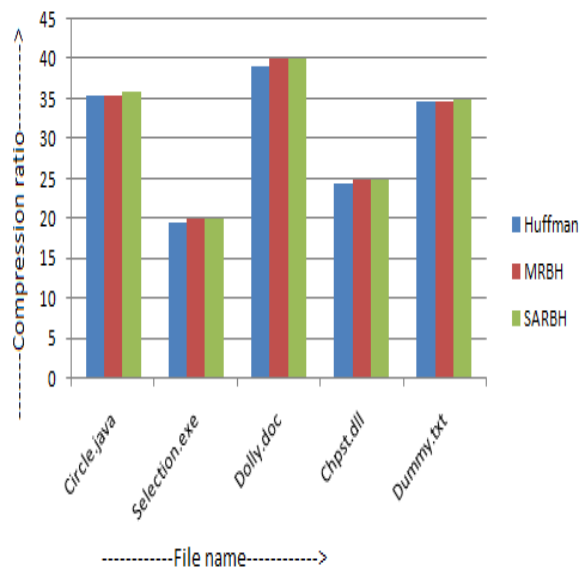

Fig .4. Graphical representation of compression ratios of Huffman, MRBH and SARBH coding.

## IV. CONCLUSION

The proposed technique eliminates some of the limitations of classical Huffman, RBH and MRBH coding techniques and offers better performance for some files over both Huffman and MRBH coding. The presented technique has also a better scope of modification. The technique can be modified by introducing the concept of region wise code interchanging. And it can also be used for image compression.

### ACKNOWLEDGMENT

The authors extend sincere thanks to the department of Computer Science and Engineering and IIPC Cell, University of Kalyani, Nadia, West Bengal, India for using the infrastructure facilities for developing the technique.